\documentstyle[12pt]{article}
\begin{document}
\textheight 9in
\topmargin -.5in
\vbadness = 100000
\hbadness = 100000
\title{The lack of rotation in the Trouton-Noble experiment}
\author{Jerrold Franklin\footnote{Internet address:
Jerry.F@TEMPLE.EDU}\\
Department of Physics\\
Temple University, Philadelphia, PA 19122-6082}
\maketitle
\begin{abstract}
The absence of any tendency toward rotation in the Trouton-Noble experiment
is given a simple explanation.  
\end{abstract}

The Trouton-Noble (TN) experiment\cite{tn,janssen} was performed in 1903 in an attempt to measure the velocity of the Earth's movement through the ``aether" by observing the rotation of a charged capacitor.  The belief was that the electric field between the plates of the capacitor would lead to a magnetic field in a moving capacitor that would produce a rotation of the capacitor.  In the actual experiment, no rotation of the moving capacitor was observed.  The ``Trouton-Noble paradox" arises because simple calculation shows that ${\bf r\times}({\bf dp}/dt)$ on a moving capacitor plate is non-zero.  If ${\bf r\times}({\bf dp}/dt)$ is interpreted as the ``torque", that leads to the question ``Why doesn't a moving capacitor rotate?"  The TN experiment seemed to pose a problem for special relativity because the relativistic calculation of the cross product 
${\bf r\times}({\bf dp}/dt)$ on a moving capacitor plate is still non-zero.  
In this paper, we give a simple explanation of why, in spite of this, there is no tendency for the moving capacitor to rotate.   

Although the TN experiment is a clear and simple example of the failure of non-relativistic theory, it is not found in many textbooks.
Perhaps this is because there has been no simple (and correct) explanation for the lack of rotation of a moving capacitor, even in special relativity.  The experiment is described by Panofsky and Phillips,\cite{pp} and the apparent ``paradox" is discussed there.   They make the unfortunate statement (on p. 349): ``The torque predicted here is real enough to an observer moving with a velocity {\bf u} relative to the two charges, and should in that case be measurable if there were no mechanical considerations involved."  Further incorrect statements follow about ``elastic stresses" depending on the velocity of the moving observer.  

There is a large number of papers addressing the question of why a moving capacitor does not rotate.\cite{refs}  They generally try to show that ${\bf r\times}({\bf dp}/dt)$ does become zero due to cancellations between the Lorentz force on a capacitor plate and either the momentum change of the EM field or stress forces in the capacitor plate.  
We show in the appendix of this paper that both of those approaches are incorrect.    

The null result of the TN experiment was briefly discussed by the present author in \cite{jf}, where the apparent "paradox" was shown to be due to a misinterpretation of the physical meaning of ${\bf r\times}({\bf dp}/dt)$ in relativity. The present paper develops this and extends it to treat the parallel plate capacitor case explicitly as well as some non-electromagnetic phenomena.

We consider first the case of two point charges, $+q$ and $-q$, connected by a rigid rod moving with velocity $\bf v$.
The electromagnetic field for a point charge $q$ moving with constant velocity $\bf v$ is given by
\begin{eqnarray}
{\bf E}&=&\frac{q{\bf r}}{\gamma^2\left[{\bf r}^2
-{(\bf v\times r})^2\right]^\frac{3}{2}}\\
{\bf B}&=&{\bf v\times E}, 
\end{eqnarray}
where $\gamma=1/\sqrt{1-{\bf v}^2}$.
Then, the rate of change of momentum of either particle at the end of the moving rod (if free to accelerate) is given by the Lorentz force
\begin{equation}
\frac{d{\bf p}}{dt}=-q({\bf E+v\times B})
=\frac{-q^2[{\bf r}+{\bf v\times(v\times r)}]}
{\gamma^2[{\bf r}^2-({\bf v\times r)^2}]^{\frac{3}{2}}},
\label{eq:dpdt}
\end{equation}
where $\bf r$ is the vector distance from the other charge.
We use the Gaussian system, and choose units such that $c=1$. 

The fact that ${\bf r}\times\frac{\bf dp}{dt}\ne 0$ (even in this relativistic calculation) would seem to indicate that the charges would rotate.  However, if we investigate the tendency of the charges to rotate about their center of mass, we find that there is no tendency to rotate.  This can be seen by looking at the relativistic connection between $\frac{\bf dp}{dt}$ and acceleration $\bf a$ (defined as the rate of change of velocity):
\begin{equation}
\frac{d\bf p}{dt}=\frac{d}{dt}(m{\bf v}\gamma)
= m\frac{d}{dt}\left[\frac{\bf v}
{\sqrt{1-{\bf v}^2}}\right]
=m\gamma^3[{\bf a}+{\bf v\times(v\times a)}].
\label{eq:acc}
\end{equation}
Combining Eqs. (\ref{eq:dpdt}) and (\ref{eq:acc}) gives, for the initial acceleration of either particle,
\begin{equation}
m\gamma^3[{\bf a}+{\bf v\times(v\times a)}]
=\frac{-q^2[{\bf r}+{\bf v\times(v\times r)}]}
{\gamma^2[{\bf r}^2-({\bf v\times r)^2}]^{\frac{3}{2}}}.
\label{eq:acc2}
\end{equation}

This looks like a complicated equation for $\bf a$, but it has the relatively simple solution 
(by comparison of the numerator of the RHS with the LHS)
\begin{equation}
{\bf a}=\frac{-q^2{\bf r}}
{m\gamma^5[{\bf r}^2-({\bf v\times r)^2}]^{\frac{3}{2}}}.
\label{eq:not}
\end{equation}
This result shows that the initial acceleration of either particle would be in the same direction as $\bf r$, directly toward the other particle.  This means that there would be no tendency for the moving rod in this example to rotate. Even though ${\bf r\times(dp}/dt)$ is non-zero, ${\bf r\times a}$ does equal zero and there is no rotation.
Although ${\bf dp}/dt$ is directly related to the EM fields by the Lorentz force equation, it is the acceleration $\bf a$ that determines how a particle would move. 
We see that if a pre-relativistic definition of ``torque" 
as ${\bf r\times(dp}/dt)$ is used, there could be a torque, but no tendency to rotate.  If the problem is discussed in terms of ``tendency to rotate" there is no paradox to explain.  It is only if a definition of ``torque" taken over from pre-relativistic physics is used that confusion enters.

J. D. Jackson\cite{jdj} has treated a somewhat similar case in which one of the charges is free to move.  He shows that, although the free charge follows a curved path in the moving frame (found by Lorentz transforming its trajectory),  the vector between the charges retains a fixed direction with respect to the velocity of the moving frame.
This is in agreement with our result that there is no tendency for rotation
even though ${\bf r\times(dp}/dt)\ne 0$ in the moving frame.

We have so far considered the case of two point charges, while the TN experiment was for a moving charged parallel plate capacitor.  For the case of the Trouton-Noble moving parallel plate capacitor, we consider the force on one plate due to the other plate.  The electric field  inside a parallel plate capacitor at rest is given by
\begin{equation}
{\bf E'}=\frac{E_0{\bf d'}}{d'},
\end{equation}
where $\bf d'$ is a vector from the middle of the positive plate to the middle of the negative plate, and $E_0$ is the constant magnitude of $\bf E'$.  There is no magnetic field for the stationary capacitor.
The electromagnetic fields for a capacitor moving with velocity $\bf v$ can be found by a Lorentz transformation to a system moving with velocity 
$\bf -v$.  The resulting fields are  
\begin{eqnarray}
{\bf E}&=&\frac{E_0{\bf d}}{\sqrt{{\bf d}^2-{(\bf v\times d})^2}}\\
{\bf B}&=&{\bf v\times E}
=\frac{E_0{(\bf v\times d)}}{\sqrt{{\bf d}^2-({\bf v\times d)}^2}}, 
\end{eqnarray}
where $E_0$ is still the magnitude of the electric field in the stationary capacitor, while $\bf d$ is the vector from the middle of the positive plate to the middle of the negative plate in the moving system.

The force on one plate due to the other plate is given by the Lorentz force equation
\begin{equation}
\frac{d{\bf p}}{dt}=-\frac{1}{2}Q({\bf E+v\times B})
=\frac{-QE_0[{\bf d+v\times(v\times d)}]}
{2\sqrt{{\bf d}^2-{(\bf v\times d})^2}},
\end{equation}
where $-Q$ is the charge on the negative plate, and the factor of 
$\frac{1}{2}$ enters because the fields between the plates are due to both plates.  Again using Eq. (\ref{eq:acc}) for the acceleration (if the plates were free to accelerate), we get 
\begin{equation}
m\gamma^3[{\bf a}+{\bf v\times(v\times a)}]
=\frac{-QE_0[{\bf d+v\times(v\times d}]}
{2\sqrt{{\bf d}^2-{(\bf v\times d})^2}}.
\end{equation}  
As before, this has the solution
\begin{equation}
{\bf a}=\frac{-QE_0{\bf d}}
{2m\gamma^3\sqrt{{\bf d}^2-{(\bf v\times d})^2}}.
\end{equation}
This means that the tendency for acceleration of one plate (if let loose) is directly toward the other plate,
and there is no tendency for the capacitor to rotate.

We note that, in both cases, the relativistic transformation of the electromagnetic fields leads to a factor [$\bf r+ v\times(v\times r)$]
in the Lorentz force between two charges, so
${\bf r\times(dp}/dt)$ is non-zero.
However, a similar factor [$\bf a+ v\times(v\times a)$], appears in the relativistic connection between ${\bf dp}/{dt}$ and the acceleration $\bf a$.  Thus, even though ${\bf r\times(dp}/dt)$ is non-zero,
there is no tendency for rotation of the two charges.

The fact that a Lorentz transformation on an object for which    
${\bf r\times(dp}/dt)={\bf 0}$ in its rest system can lead to 
${\bf r\times(dp}/dt)\ne{\bf 0}$ in a system where the object has velocity $\bf v$ is a general result, not limited to electromagnetic forces.  We consider a force $\bf F'$ (which we define as 
${\bf F'=dp'}/dt'$) acting at one end of
a rod of length $\bf r'$ that is at rest in system S$'$,
with $\bf r'\times F'=0$.
Any force $\bf F$ is related to the relativistic four-vector ``Minkowski force" $\cal F^\mu$ by ${\cal F}^\mu=(\gamma{\bf v\cdot F;\gamma F})$.
Thus the Minkowski force on the stationary rod is 
${\cal F'}^\mu=(0;{\bf F'})$.
The force on the rod in a system S where it is moving with velocity 
$\bf v$ is given by a Lorentz transformation of $\cal F'^\mu$
resulting in
\begin{eqnarray}
{\bf F_\parallel}&=&{\bf F'_\parallel}\\
{\bf F_\perp}&=&{\bf F'_\perp}/\gamma,
\end{eqnarray}
where the subscripts $\parallel$ and $\perp$ refer to the components parallel to and perpendicular to $\bf v$.  
The moment arm (taken as the vector from one end of the rod to the other)
transforms as
\begin{eqnarray}
{\bf r_\parallel}&=&{\bf r'_\parallel}/\gamma\\
{\bf r_\perp}&=&{\bf r'_\perp}.
\end{eqnarray} 
Thus $\bf r'\times F'$ transforms as
\begin{eqnarray}
{\bf r\times F}&=&{\bf r_\parallel\times F_\perp}
+{\bf r_\perp\times F_\parallel}
+{\bf r_\perp\times F_\perp}\nonumber\\
&=& \left(\frac{\bf r'_\parallel}{\gamma}\right)
\times\left(\frac{\bf F'_\perp}{\gamma}\right)
+{\bf r'_\perp}\times\bf F'_\parallel
+{\bf r'_\perp}
\times\left(\frac{\bf F'_\perp}{\gamma}\right)\nonumber\\
&=&(1-{\bf v}^2){\bf r'_\parallel}\times{\bf F'_\perp}
+{\bf r'_\perp}\times{\bf F'_\parallel}
+{\bf r'_\perp}\times{\bf F'_\perp}/\gamma\nonumber\\
&=& ({\bf r'\times F'})_\perp
+({\bf r'\times F'})_\parallel/\gamma
-{\bf (v\cdot r')(v\times F')}.
\end{eqnarray}
If $(\bf r'\times F')=0$, its parallel and perpendicular components must each equal zero separately, which leaves 
\begin{equation}
{\bf r\times F}=-{\bf (v\cdot r')(v\times F')}.
\end{equation}

We see that the product $(\bf r\times F)$ will not vanish for the moving rod.  However, we now show as a general result that if a force does not cause rotation in one Lorentz system (That is, $\bf r'\times a'=0$.), it will not cause rotation in any Lorentz system.  
We first show that if $\bf r'\times a'=0$ in the rest system S$'$, where $\bf r'$ is the moment arm, then $\bf r\times a=0$ in a Lorentz system S moving with velocity $\bf -v$ with respect to system S$'$.
If S$'$ is the rest system, then the transformation equations for acceleration are
\begin{eqnarray}
{\bf a_\parallel}&=&{\bf a'_\parallel}/\gamma^3\\
{\bf a_\perp}&=&{\bf a'_\perp}/\gamma^2,
\end{eqnarray}
The ``turning moment" (that is the vector that is proportional to the angular acceleration) in system S is given by
\begin{eqnarray}
{\bf r\times a}&=&{\bf r_\parallel\times a_\perp}
+{\bf r_\perp\times a_\parallel}
+{\bf r_\perp\times a_\perp}\nonumber\\
&=& \left(\frac{\bf r'_\parallel}{\gamma}\right)
\times\left(\frac{\bf a'_\perp}{\gamma^2}\right)
+\left({\bf r'_\perp}\right)
\times\left(\frac{\bf a'_\parallel}{\gamma^3}\right)
+\left({\bf r'_\perp}\right)
\times\left(\frac{\bf a'_\perp}{\gamma^2}\right)\nonumber\\
&=& \frac{\left({\bf r'\times a'}\right){\bf_\perp}}{\gamma^3}
+\frac{\left({\bf r'\times a'}\right){\bf_\parallel}}{\gamma^2}
={\bf 0}.
\end{eqnarray}
Thus, if ($\bf r'\times a')=0$, ($\bf r\times a$) must also vanish.
We can extend this result for any two Lorentz systems by comparing each to the rest system.  The general result is that if an object doesn't rotate in one Lorentz system, it won't rotate in any Lorentz system,
in agreement with the general principles of special relativity.
This is true even though we have also shown that ${\bf r\times (dp}/dt$) may not vanish.\\

I see two lessons from the results in this paper:
\begin{enumerate}
\item In applying special relativity, we must be careful if we use 
pre-relativistic definitions and terminology.  As examples, the use of terminology like ``force" and ``torque" can lead to error unless there is  careful definition of precisely what is meant in the context of special relativity.

\item We should treat any problem in the Lorentz system where it is simplest (e.g., the rest system of the capacitor), and rely on the Lorentz invariance of special relativity to preserve the physics for any other Lorentz system.  Doing a simple problem in an awkward Lorentz system can lead to mathematical complexity with no better understanding of the physics, and open the door to confusion.  It is obvious that the Trouton-Noble capacitor does not rotate in its rest system, but hundreds of pages 
have been written for the moving system.
\end{enumerate}
\vspace{.2in}
{\large{\bf Appendix}}\\
\\
In this appendix we investigate claims that ${\bf r\times}({\bf dp}/dt)$    of the capacitor vanishes, either because:
\begin{enumerate}
\item  ${\bf r\times}({\bf dp}/dt)$ of the capacitor is cancelled by an equal, but opposite ${\bf r\times}({\bf dp}/dt)$ of the EM field, or
\item ${\bf r\times}({\bf dp}/dt)$ of the capacitor is cancelled by an equal, but opposite ${\bf r\times}({\bf dp}/dt)$ due to stress forces in the capacitor plate.
\end{enumerate}
Some papers use energy instead of momentum to achieve the same type of
cancellation.  Incidentally, if both arguments (1) and (2) were correct, the Lorentz force would be cancelled twice.

Papers that use case (1) calculate ${\bf r\times}({\bf dp}/dt)$ of the EM field and find, not surprisingly, that it is equal and opposite to 
${\bf r\times}({\bf dp}/dt)$ of the capacitor.  This is not surprising because it is shown as a general result in all EM textbooks that these two momentum changes will always cancel.  In fact this cancellation is used in the textbooks to define the momentum ${\bf p}_{\rm EM}$ (and angular momentum ${\bf L}_{\rm EM}$) of the EM field in order that ${\bf p}_{\rm matter}+{\bf p}_{\rm EM}$ 
(and ${\bf L}_{\rm matter}+{\bf L}_{\rm EM}$) will be constant in time.  To add the two time derivatives ${\bf r\times}({\bf dp}/dt)$ as if they both act on the capacitor is wrong.  The same argument could be used to prove that an EM field can't move anything.
The argument that the change in EM field momentum cancels the Lorentz force
is also refuted by our demonstration that $\bf r\times F$ does not vanish for a general, not necessarily electromagnetic, force.

Papers that use case (2) are just complicated examples of the misuse of Newton's third law applied to a horse pulling a cart.  Consider the case of the charged particles at the ends of a rod.  This group of papers shows that there will be a force (still defined by ${\bf dp}/dt)$
on the charge, exerted by the rod, that is equal and opposite to the Lorentz force exerted on the charge.  
This leads to a cancellation which these papers say prevent the rod from rotating.  But, as with the horse and the cart, there will be a reaction force of the charged particle on the rod that is not in general along the length of the rod.  This force would act to rotate the rod, except for the effect we have shown in this paper.  Although we have used the word
``force" to denote ${\bf dp}/dt$, the tendency for motion would not be in the direction of this ``force".  

Another way to see that stress in the rod cannot explain the TN experiment is to consider the case of two positive charges connected by a flexible string.  The reaction ${\bf dp}/dt$ on the moving string will not be along the string, but the stress tension in the string can only be along the string.  The direction of the tension is determined by the direction the charges would move if the string were cut, which we have shown is along the string, so the moving string would not rotate even though 
${\bf r\times(dp}/dt)\ne{\bf 0}$.

It is also interesting to note that both approaches (1) and (2) need no aspect of special relativity to keep the capacitor from rotating.  In fact,  
they derive from Lorentz's early attempt to avoid relativity in explaining the null result.
In our derivation, the key input is Eq. (\ref{eq:acc}) giving the relativistic connection between acceleration (how something moves) and 
${\bf dp}/dt$ (how it interacts), showing that they are not always in the same direction.  
Thus, understanding the Trouton-Noble experiment does require special relativity, in spite of Lorentz's (and later) attempts to explain it away.

\end{document}